\catcode`\@=11
 
\def\citen#1{\if@filesw \immediate\write \@auxout {\string\citation{#1}}\fi%
\@tempcntb\m@ne \let\@h@ld\relax \def\@citea{}%
\@for \@citeb:=#1\do {\@ifundefined {b@\@citeb}%
    {\@h@ld\@citea\@tempcntb\m@ne{\bf ?}%
    \@warning {Citation `\@citeb ' on page \thepage \space undefined}}%
    {\@tempcnta\@tempcntb \advance\@tempcnta\@ne
    \setbox\z@\hbox\bgroup\ifcat0\csname b@\@citeb \endcsname \relax
    \egroup \@tempcntb\number\csname b@\@citeb \endcsname \relax
    \else \egroup \@tempcntb\m@ne \fi \ifnum\@tempcnta=\@tempcntb
    \ifx\@h@ld\relax \edef \@h@ld{\@citea\csname b@\@citeb\endcsname}%
    \else \edef\@h@ld{\hbox{--}\penalty\@highpenalty
    \csname b@\@citeb\endcsname}\fi
    \else \@h@ld\@citea\csname b@\@citeb \endcsname \let\@h@ld\relax \fi}%
\def\@citea{,\penalty\@highpenalty\hskip.13em plus.13em minus.13em}}\@h@ld}
\def\@citex[#1]#2{\@cite{\citen{#2}}{#1}}%
\def\@cite#1#2{\leavevmode\unskip\ifnum\lastpenalty=\z@\penalty\@highpenalty\fi%
  \ [{\multiply\@highpenalty 3 #1%
  \if@tempswa,\penalty\@highpenalty\ #2\fi}]}   %
\makeatother 

\catcode`\@=12

\newcommand\barray[6]{\!{\scs\begin{array}{ccc}{}\\[-1.94em]
                   \!{\scs #1}\!&\!{\scs #2}\!&\!{\scs #3}\!\\[-.43em]\!{\scs #4}
                   \!&\!{\scs #5}\!&\!{\scs #6}\! \\[-.4em] \end{array}}\!}

\def\Bc            {Boundary condition}
\def\be            {\begin{equation}}
\def\bearl         {\begin{array}{l}}

\def\cft           {conformal field theory}

\def\con           {conformal }
\def\corfu         {correlation function}
\def\CS            {Chern\hy Si\-mons }

\def\dim           {dimension}
\def\disc          {disk}
\def\dl            {\mathbb }
\def\dsty          {\displaystyle}

\def\ee            {\end{equation}}
\def\eE            {{\rm e}}
\def\eear          {\end{array}}
\def\eP              {\end{picture}}

\def\eq            {\,{=}\,}
\newcommand\erf[1] {(\ref{#1})}

\newcommand\Frac[2]{\mbox{\large$\frac{#1}{#2}$}}
\def\furu          {fusion rule}
\def\futnote#1     {\footnote{~#1}\ }

\def\hy            {$\mbox{-\hspace{-.66 mm}-}$}

\def\ii            {{\rm i}}
\def\iN            {\,{\in}\,}

\long\def\labl#1   {\label{#1}\ee}

\def\llb           {\mbox{\large(}}
\def\lrb           {\mbox{\large)}}

\def\ot            {\raisebox{.07em}{$\scriptstyle\otimes$}}
\def\oT            {\,\ot\,}
\def\parfu         {partition function}
\def\preimage      {pre-ima\-ge}
\def\preimages     {pre-ima\-ges}
\def\Q             {Quantum }
\def\qft           {quantum field theory}

\def\rpd           {\mbox{$\dl{RP}^3$}}
\def\rpz           {\mbox{$\dl{RP}^2$}}
\def\scs           {\scriptstyle}
\newcommand\sixj[8]{\mbox{\Large\{} \barray{#1}{#2}{#3}{#4}{#5}{#6}
                   \mbox{\Large\}} ^{\!#7}_{\!#8} }

\def\sym           {symmetry}

\def\twodim        {two-dimensional}

\def\Wedge         {\,{\wedge}\,}

\def\wzwm          {WZW model}

\def\zet           {{\dl Z}}

     \documentclass[12pt]{article}
     \usepackage{amssymb,amsfonts} \usepackage[dvips]{graphics}
\begin{document}


  \begin{flushright}  {~} \\[-1cm] {\sf hep-th/9909140} \\[1mm]
  {\sf ETH-TH/99-25} \\[1 mm]
  {\sf ESI-759} \\[1 mm]
  {\sf September 1999} \end{flushright}
 
  \begin{center} \vskip 22mm
  {\Large\bf CONFORMAL BOUNDARY CONDITIONS}\\[1.02em]
  {\Large\bf AND THREE-DIMENSIONAL}\\[1.02em]
  {\Large\bf TOPOLOGICAL FIELD THEORY}\\[22mm]
  {\large Giovanni Felder}\,, \ {\large J\"urg Fr\"ohlich}\,, \ 
  {\large J\"urgen Fuchs} \ and \ {\large Christoph Schweigert}\\[5mm]
  ETH Z\"urich\\[.2em] CH -- 8093~~Z\"urich
  \end{center} \vskip 26mm
  \begin{quote}{\bf Abstract}\\[1mm]
We present a general construction of all correlation functions of a 
two-dimensional rational conformal field theory, for an arbitrary number 
of bulk and boundary fields and arbitrary topologies. The correlators are 
expressed in terms of Wilson graphs in a certain three-manifold, the 
connecting manifold. The amplitudes constructed this way can be shown to 
be modular invariant and to obey the correct factorization rules.
  \end{quote} \newpage


Two-dimensional conformal field theory plays a fundamental role in the theory
of two-dimensional critical systems of classical statistical mechanics
\cite{frqs2}, in quasi one-dimensional condensed matter physics \cite{affl6}
and in string theory \cite{bepz}. The study of defects in systems of
condensed matter physics \cite{osaf}, of percolation probabilities \cite{card12}
and of (open) string perturbation theory in the background of certain string
solitons, the so-called D-branes \cite{polc3}, forces one to analyze
conformal field theories on surfaces that may have boundaries and\,/\,or
can be non-orientable.

In this letter we present a new description of correlation functions
of an arbitrary number of bulk and boundary fields on general surfaces.
We also show how to compute various types of operator product
coefficients from our formulas. For simplicity, 
in this letter we restrict our attention to boundary conditions that 
preserve all bulk symmetries. Moreover, we take the modular invariant torus
partition function that encodes the spectrum of bulk fields of the theory
to be given by charge conjugation.
Technical details and complete proofs will appear in a separate publication. 

Given a chiral \cft, such as a chiral free boson,
our aim is to compute correlation functions on a \twodim\ surface $X$ 
that may be non-orientable and can have a boundary. To this end, we first 
construct the so-called {\em double\/} $\hat X$ of the surface $X$. This
is an oriented surface, on which an orientation
reversing map $\sigma$ of order two acts in such a way that $X$ is obtained as
the quotient of $\hat X$ by $\sigma$. Thus $\hat X$ is a two-fold cover
of $X$; but this cover is branched over the boundary points, which correspond 
to fixed points of the map $\sigma$. For example, when $X$ is the \disc\ $D$,
then $\hat X$ is the two-sphere and $\sigma$ is the reflection about its 
equatorial plane. For $X$ the cross-cap, i.e.\ the real projective plane \rpz, 
$\hat X$ is
again the two-sphere, but $\sigma$ is now the antipodal map. Finally, when
$X$ is closed and orientable, the double  $\hat X$ consists of two disconnected
copies of $X$ with opposite orientation, $\hat X\,{\cong}\,X{\sqcup}(-X)$.

Quite generally, correlation functions on a surface $X$ can be constructed 
from conformal blocks on its double $\hat X$ \cite{ales,fuSc6}. As a first step,
one has to find the \preimages\ on $\hat X$ of all insertion points on $X$, and 
associate a primary field
of the chiral \cft\ to each of them. Since bulk points have two \preimages, 
for a bulk field two chiral labels $j$ and $j^*$ are needed, corresponding to 
left and right movers. Boundary fields, in contrast, carry a single label
$k$; yet, they should {\em not\/} be thought of as chiral objects.

Having associated these labels to the geometric data, we can assign
a vector space of conformal blocks, not necessarily
of dimension one, to every collection of bulk and boundary fields on $X$.
The correlation function is one specific element in this space.
This element must obey modular invariance and factorization properties. 
The conformal bootstrap programme \cite{bepz} allows to determine the 
correlation function by imposing these properties as constraints.  
Fortunately, the connection between conformal field theory in two dimensions
and topological field theory in three dimensions supplies us with a most direct
way to construct concrete elements in the spaces of conformal blocks. 
What one must do in order to specify a a definite element in the space
of conformal blocks is to find a three-manifold $M_X$ whose boundary is $\hat X$,
  \be  \partial M_X = \hat X \,,  \ee
as well as a Wilson graph $W$ in $M_X$ that ends at the marked points on 
$\hat X$. This can be done for any arbitrary rational \cft; for details, 
which are based on the axiomatization in \cite{TUra}, we refer to \cite{fffs3}.
In the particular case of \wzwm s, Chern\hy Simons theory can be used 
\cite{witt27,frki2,hora5} for this construction. For these models, the element 
in the space of conformal blocks is obtained by the Chern\hy Simons path integral
  \be  \int\!\! {\cal D}\!A\  W\,
  \exp\llb\, \ii\,\Frac k{4\pi}\int_{\!M_X^{}}
  \!\!\!{\rm Tr}\, (A\Wedge{\rm d}A+\Frac23\,A\Wedge A\Wedge A)\,\lrb  
  \ee
with appropriate parabolic conditions at the punctures.

\smallskip
Thus to obtain a correlation function on $X$, we first construct
a certain three-manifold $M_X$ with boundary $\hat X$, which we call the 
{\em connecting three-manifold\/}. Technically, the manifold
$M_X$ can be characterized as follows. When $X$ does not have a boundary, 
then $M_X\eq(\hat X{\times}[-1{,}1])/\zet_2$, where
the group $\zet_2$ acts on $\hat X$ by $\sigma$ and on the interval $[-1{,}1]$
by the sign flip $t\,{\mapsto}\,{-}t$ for $t\iN[-1{,}1]$. Thus $M_X$ consists 
of pairs $(x,t)$ with $x$ a point on the double $\hat X$ and $t$ in 
$[-1{,}1]$, modulo the identification $(x,t)\,{\sim}\,(\sigma(x),-t)$. 
For fixed $x$, the points of the form $(x,t)$ form a segment, 
the {\em connecting interval\/}, joining the two \preimages\ of a point 
in $X$. When $X$ has a boundary, we obtain $M_X$ from 
$(\hat X\,{\times}\,[-1{,}1])/\zet_2$ by contracting the connecting intervals 
over the boundary to single points, in such a way that $M_X$ remains a smooth 
manifold.
(An equivalent construction, in which the boundary intervals are not
contracted, was given in \cite{hora5}.)

It is readily checked that the boundary of the connecting manifold $M_X$ is 
indeed the double $\hat X$. Moreover, $M_X$
connects the two \preimages\ of a bulk point by an interval in such
a manner that the connecting intervals for distinct bulk points do not 
intersect. Let us list
a few examples. For a \disc, the connecting manifold is a solid three-ball, 
and the connecting intervals are all perpendicular to the equatorial plane.
Similarly, when $X$ is the annulus, $M_X$ is a solid torus.
For $X$ the cross-cap, the connecting manifold $M_X$ is best characterized 
by the fact that when glueing to its boundary a solid ball, we obtain 
$S^3/\zet_2\,{\cong}\,\rpd$, which coincides with the group manifold
of the Lie group SO$(3)$. For closed orientable surfaces $X$, the bundle 
$M_X$ is just the trivial bundle $X\,{\times}\,[-1{,}1]$; e.g.\ when $X$ is a 
sphere, then $M_X$ can be visualized as consisting of the points between 
two concentric spheres.

\begin{figure}[htbp] \centering
  \scalebox{0.28}{\includegraphics{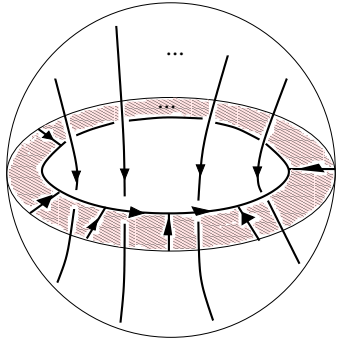}}
  \caption{Wilson graph for the \disc\ correlators}
\label{m1}\end{figure}

The next step is to specify a certain Wilson graph in $M_X$. The prescription, 
which is illustrated in figure \ref{m1} for the case of a \disc\ with an 
arbitrary number of insertions in the bulk and on the boundary, is as follows. 
First, for every bulk insertion $j$, one joins the \preimages\ of the
insertion point by a Wilson line running along the connecting interval. Next, 
one inserts one circular Wilson line parallel to each component of the 
boundary (a similar idea was presented in \cite{hora5})
and joins every boundary insertion $k$ on the respective boundary 
component by a short Wilson line to the corresponding circular Wilson line.
Moreover, the circular Wilson lines are required to
run ``close to the boundary", in the sense that none of the connecting
intervals of the bulk fields passes between the circular Wilson lines
and the boundary of $X$.

So far we have only specified the geometric information for the conformal
blocks. To proceed, we also must attach a primary label of the chiral \cft\ 
to each segment of the Wilson graph. For the bulk points, this 
prescription is immediate, as we are dealing with the charge conjugation
modular invariant. Similarly, we are naturally provided with the labels $k$ for 
the short Wilson lines that connect the boundary insertions with the circular 
Wilson lines. In addition, the segments of the circular Wilson lines should
encode the boundary conditions of the corresponding boundary segments. Recalling
that those boundary conditions which preserve all bulk symmetries can be 
labelled by the primary fields of the chiral \cft\ \cite{card9}, we 
attach such a primary label $a$ to every segment of the circular 
Wilson lines. Finally, we must consider the three-valent junctions on the 
circular Wilson lines. For each of them we choose an element $\alpha$
in the space of chiral couplings between the label $k$ for the boundary field
and the two adjacent boundary conditions $a,\,b$. The dimension of
this space of couplings is given by the fusion rules ${\rm N}_{kb}^{\;\ a}$
of the chiral theory.
Indeed, it is known that boundary operators need an additional degeneracy 
label that takes its values in the space of chiral three-point blocks.

As a matter of fact, every segment of the Wilson graph should also be equipped
with a framing \cite{witt27} -- in other words, we should not just specify a 
graph, but a ribbon graph. 
Moreover, the boundary $\hat X$ of $M_X$ must be endowed with additional 
structure, too. A careful discussion of these issues will be presented 
in \cite{fffs3}. As a side remark, we mention that the circular 
Wilson lines already come with a natural thickening to ribbons, which is
obtained by connecting them to the \preimage\ of the boundary of $X$ in
$\hat X$. (In figure \ref{m1} this is indicated by a shading.) 
Note that in the case of symmetry breaking boundary conditions \cite{fuSc1112}
the labels of boundary fields and boundary conditions can be
more general than in the bulk. This can be implemented in our picture, as the 
corresponding part of the graph with the circular Wilson line is disconnected 
from the rest of the Wilson graph.

\smallskip
Using appropriate surgery on three-manifolds, we can prove that
the correlation functions obtained by our prescription possess the
correct factorization (or sewing) properties and that they are invariant 
under large diffeomorphisms or, in more technical terms, under the
relative modular group \cite{bisa2}. For a detailed
account of these issues we refer to \cite{fffs3}. Here we
restrict ourselves to the analysis of a few situations of particular
interest; we also show how to 
recover known results for the structure constants from our formulas.

In our approach the structure constants are obtained as the coefficients 
that appear in the expansion of the specific element in the space of 
conformal blocks that represents a correlation function 
in a standard basis for the conformal blocks. 
For two points on the boundary of a solid three-ball such a standard basis 
is given by a Wilson line (with trivial framing) connecting the two points, 
while for three points one takes a Mercedes star shaped junction of three 
Wilson lines. Our general strategy for computing the coefficients is then
to glue another three-manifold to the connecting manifold so as to obtain 
the partition function or, in mathematical terms, the link invariant,
for a closed three-manifold. The values of such link invariants are available
in the literature, see e.g.\ \cite{witt27,frki2,TUra,Mose}.

\begin{figure}[htbp] \centering
  \scalebox{0.2}{\includegraphics{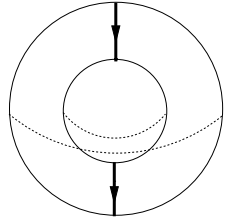}}
  \caption{$C(S^2;j,j^*)$}
  \label{m2} \end{figure}

\begin{figure}[htbp] \centering
  \scalebox{0.2}{\includegraphics{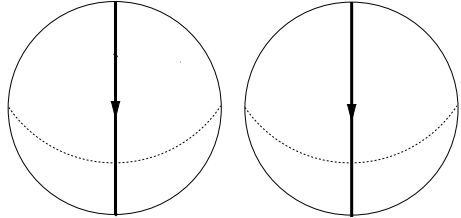}}
  \caption{$B(S^2;j,j^*) \,{\scriptstyle\otimes}\, B(-S^2;j,j^*)$}
  \label{m3} \end{figure}

\smallskip
Our first example is the correlator of two (bulk) fields on $S^2$, 
a closed and orientable surface. For the space of blocks to be non-zero, the
two fields must be conjugate, i.e.\ carry labels $j$ and $j^*$, respectively.
According to our prescription, the connecting
manifold then consists of the filling between two concentric two-spheres, and 
the Wilson graph consists of two disjoint lines connecting the spheres, both
labelled by $j$; this is depicted in figure \ref{m2}. The space of conformal 
blocks for this situation is one-dimensional; its standard basis 
is displayed in figure \ref{m3}. Thus the relevant three-manifold is given by the
disconnected sum of two balls, each of which carries a single Wilson line. 
To both manifolds we glue two balls in which a Wilson line labelled by $j$
is running. In the case of the \corfu, the resulting manifold is a three-sphere 
with an unknot
labelled by $j$, for which the value of the link invariant is $S_{0,j}$.
($S$ is the modular S-transformation matrix of the chiral conformal field 
theory, and the label $0$ refers to the vacuum primary field.) When applied
to the manifold in figure \ref{m3}, the glueing procedure produces
two disjoint copies of $S^3$, each with
an unknot labelled by $j$; the corresponding partition function is $S_{0,j}^2$. 
Comparing the two results we see that the two-point function on the sphere is
expressed in terms of the standard basis as
  \be  C(S^2;j,j^*) = S_{0,j}^{-1} \cdot B(S^2;j,j^*) \oT B(-S^2;j,j^*)
  \,.  \labl3
In other words, the normalization of the bulk fields $j$ differs 
by a factor of $(S_{0,j})^{-1/2}$ from the more conventional prescription 
where they are `canonically normalized to one'. 

\begin{figure}[htbp] \centering
  \scalebox{0.2}{\includegraphics{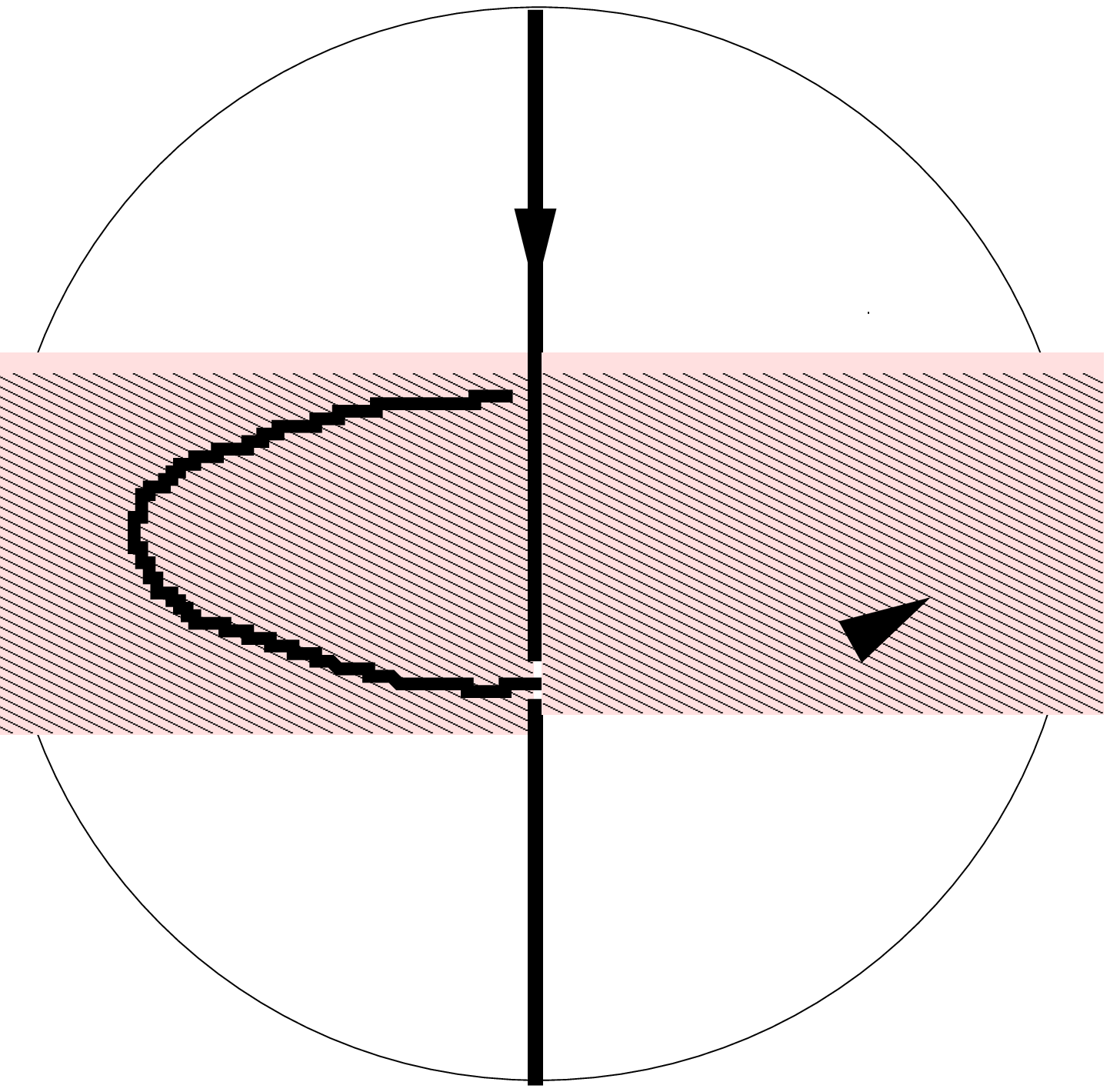}}
  \caption{$C(D_a;j)$}
  \label{m4}\end{figure}

\begin{figure}[htbp] \centering
  \scalebox{0.2}{\includegraphics{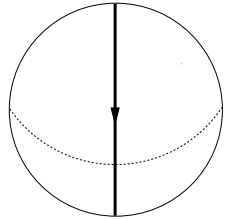}}
  \caption{$B(S^2;j,j^*)$}
  \label{m5}\end{figure}

\smallskip
Next we discuss an example featuring an orientable surface with boundary; we
compute the one-point amplitude for a bulk field $j$ on a disk $D$ with 
boundary condition $a$. Again the space of blocks is one-dimensional.
Our task is then to compare the Wilson graph of figure \ref{m4} with 
the standard basis that is displayed in figure \ref{m5}. 
(In the present context, this particular conformal block is 
often called an `Ishibashi state'). We now obtain the three-sphere by
glueing with a single three-ball. When applied to the graph of figure 
\ref{m5}, we get the unknot with label $j$ in $S^3$, for which the 
\parfu\ is $S_{0,j}$. In the case of figure \ref{m4} we get a pair of linked 
Wilson lines with labels $a$ and $j$ in $S^3$; the value of the link 
invariant for this graph 
is $S_{a,j}$. Comparison thus shows that the correlation function is
$S_{a,j}/S_{0,j}$ times the standard two-point block on the sphere,
  \be  C(D_a;j) = (S_{a,j}/S_{0,j}) \cdot B(S^2;j,j^*) \,.  \labl4
Taking into account the normalization of the bulk fields as obtained in
formula \erf3, we recover the known result that 
the correlator for a canonically normalized bulk field $j$ 
on a disk with boundary condition $a$ 
is $S_{a,j}/\!\sqrt{S_{0,j}}$ times the standard two-point block
on the sphere. (This relation forms the basis of the so-called boundary 
state formalism \cite{card9}.)

\begin{figure}[htbp] \centering
  \scalebox{0.25}{\includegraphics{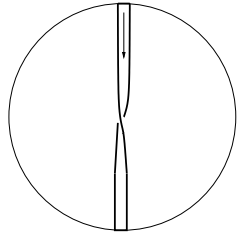}}
  \caption{The state $\psi_j^+{=}\,\eE^{-\pi\ii\Delta_j}\psi_j^{}$}
  \label{m6}\end{figure}

  \smallskip
As a third example, we study again a one-point correlator of a
(bulk) field $j$, now on the cross-cap \rpz, which does not have a
boundary, but is non-orientable. The latter property forces us to be 
careful with the framing. The structure constants are obtained by comparing 
the correlator $C(\rpz;j)$ with the `cross-cap state' $\psi_j$. This state 
is defined
in figure \ref{m6}; it is similar to the basis element $B(S^2;j,j^*)$ of
the two-point blocks on $S^2$, but now the Wilson line in the three-ball 
has a non-trivial framing, and accordingly in figure \ref{m6} we have drawn a
ribbon instead of a line. A priori we could twist the line
either by $+\pi$, thereby obtaining some state $\psi_j^+$, or by $-\pi$ and 
obtain another state $\psi_j^-$. These two vectors differ by a factor of
$\eE^{2\pi\ii\Delta_j}$, with $\Delta_j$ the conformal weight of $j$.
Salomonically, we define the cross-cap state as
  \be  \psi_j^{} := \eE^{-\pi\ii\Delta_j}\, \psi_j^- = \eE^{\pi\ii\Delta_j}\,
  \psi_j^+ \,. \ee
Again the comparison of the correlator $C(\rpz;j)$ with the standard basis
$\psi_j$ is carried out by glueing a three-ball with a Wilson line 
    to the ball of figure \ref{m6}.
In contrast to the previous cases, however, this line is given a non-trivial
framing; choosing the framing in such a way that the twist of the cross-cap
state is undone, glueing the ball to the cross-cap state yields $S^3$ 
with the unknot, with partition function $Z(S^3;j)\eq S_{0,j}$. 

\begin{figure}[htbp] \centering
  \scalebox{0.27}{\includegraphics{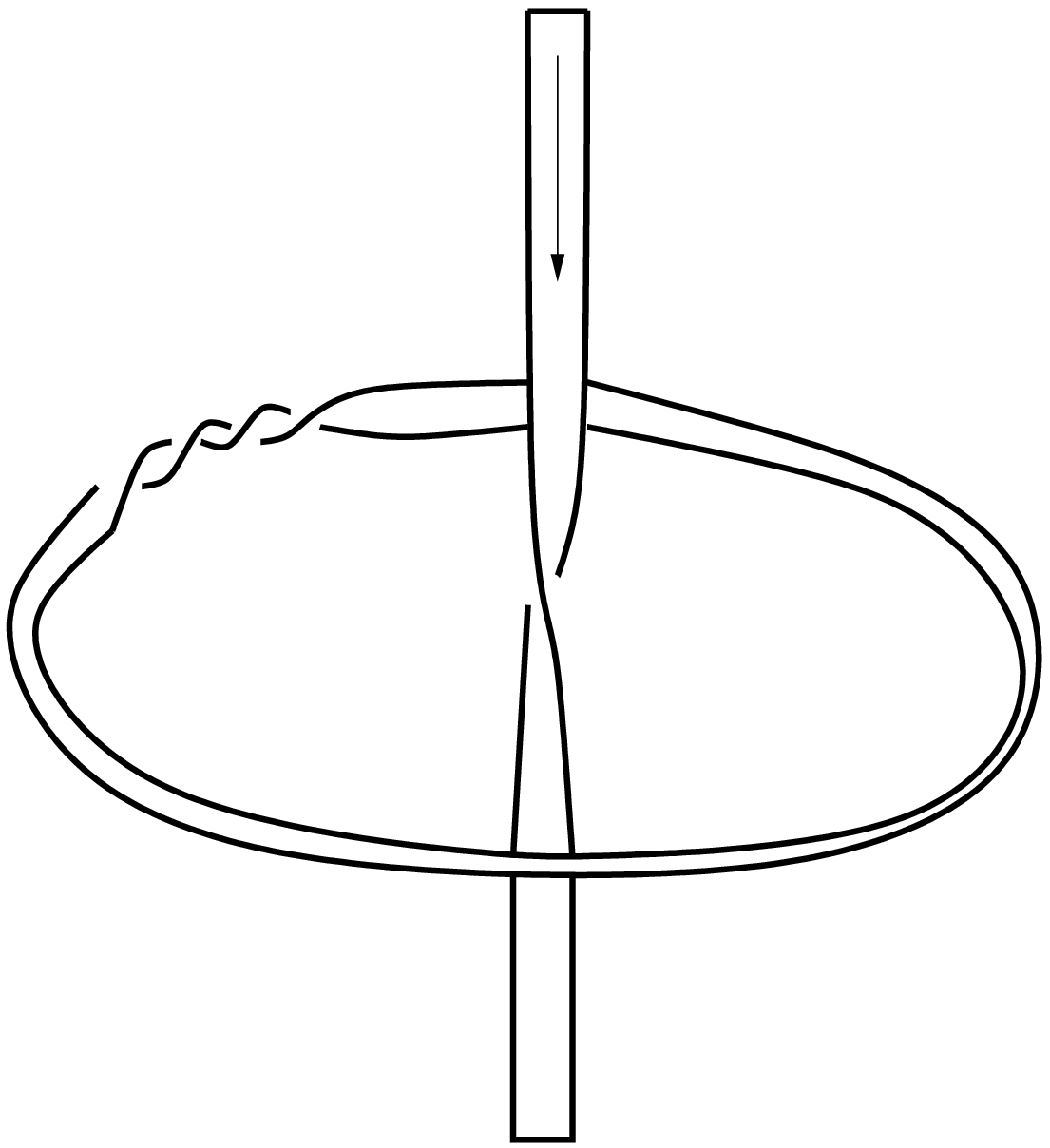}}
  \caption{A visualization of $C(\rpz;j)$}
  \label{m7}\end{figure}

As already mentioned, glueing the three-ball to the connecting manifold of 
the cross-cap yields SO$(3)$. It is also known that SO$(3)$ can be obtained 
from $S^3$ by a surgery on the unknot with framing $-2$. (Following how
the framed graph is mapped by the surgery, one may visualize the situation 
as in figure \ref{m7}.) Taking all framings properly into account, we obtain 
  \be  Z({\rm SO}(3);j) = T_0^{1/2} \sum_k S_{0,k}^{}\, (T_k^{})^2_{}\,
  S_{k,j}^{}\, T_j^{1/2} = P_{0,j}^{}  \ee
(with $T_j\,{\equiv}\,\eE^{2\pi\ii(\Delta_j-c/24)}$)
for the invariant of this three-manifold, where in the second equality we
expressed the result through the matrix \cite{bisa}
$P\,{:=}\,T^{1/2} S T^2 S T^{1/2}$. We have thereby recovered the known 
formula
  \be  C(\rpz;j) = (P_{0,j}/S_{0,j}) \cdot \psi_j \ee
for the one-point correlator on the cross-cap.
  \begin{picture}(1,1)(-233,-270) \put(0,0){{\it j}}\end{picture}
  \begin{picture}(1,1)(-187,-92) \put(0,0){{\it k}}\end{picture}
  \begin{picture}(1,1)(-225,-130) \put(0,0){{\it j}}\end{picture}

  \smallskip
As a final example, consider three boundary fields $i,\,j,\,k$ on a \disc. 
The relevant Wilson graph in the three-ball is of the type shown in figure 
\ref{m1}, without any vertical Wilson lines along connecting intervals;
it consists of a circular line (with segments labelled $a,\,b,\,c$) with 
three short Wilson lines (labelled $i,\,j,\,k$) attached to it. We must
compare it to the standard basis for three-point blocks on the sphere,
which is a Mercedes star shaped junction. This comparison can be made by 
performing a single fusing operation, followed by a contraction of the
loop. For boundary fields, it is natural to define the correlation functions 
as linear forms on the degeneracy spaces for boundary operators. Denoting a 
basis of the degeneracy space for the boundary operator $\psi^{ac}_i$ by
$\{e_\alpha[ica^*]\}$, normalized by the quantum trace condition
${\rm{tr}}\,(e_\alpha[ica^*]\,e_\beta[i^*ac^*])\eq\delta_{\alpha,\beta}$, 
we find that
  \be  C(D_{a,b,c};i,j,k) \,\llb e_\alpha[ica^*]\oT e_\beta[jab^*]\oT
  e_\gamma[kbc^*]\lrb = \dsty\sum_\kappa \Frac{S_{0,0}}
  {S_{k,0}}\,\sixj ica{b^*}{j^*}{k^*}{\alpha\beta}{\gamma\kappa}\,
  e_\kappa[kji] \,,  \ee
where the symbol $\sixj ijklmn{\alpha\beta}{\gamma\delta}$ denotes a
fusing matrix (or quantum $6j$-symbol) \cite{Mose,TUra}. 

\smallskip
One important conclusion we can draw from our results is that the construction
of \corfu s from conformal blocks can be performed in a
completely model-independent manner. All structure constants, for any
arbitrary \cft, can be expressed in terms of purely chiral data, such as
conformal weights, the modular S-matrix, fusing matrices and the like.
All specific properties of concrete models already enter at the chiral level.
Physical quantities, such as the magnetization of an open
spin chain or open string amplitudes in the background of D-branes, can be
expressed in terms of the correlators studied in this letter.

\small
  \newcommand\J[5]   { {\sl #5}, {#1} {#2} ({#3}) {#4} }
  \newcommand\Prep[2]{ {\sl #2}, preprint {#1}}
  \newcommand\inBO[7]{ {\sl #7}, in:\ {\em #1}, {#2}\ ({#3}, {#4} {#5}), p.\ {#6}}
   
\end{document}